\documentclass[11pt]{article}%

\usepackage{amssymb}
\usepackage{amsmath}
\usepackage{graphics}
\usepackage{epsfig}
\usepackage{amsfonts}
\usepackage{graphicx}%

\usepackage{subfig}

\usepackage{booktabs}

\newcommand{\eqb}{\begin{equation}}
\newcommand{\eqe}{\end{equation}}
\newcommand{\dmb}{\begin{displaymath}}
\newcommand{\dme}{\end{displaymath}}

\newcommand{\eab}{\begin{eqnarray}}
\newcommand{\eae}{\end{eqnarray}}

\newcommand{\e}{\mbox{e}}
\newcommand{\be}{\begin{equation}}
\newcommand{\ee}{\end{equation}}

\begin{document}
\begin{titlepage}
\begin{flushright}
\end{flushright}
\vspace{0.6cm}
\begin{center}
\huge{Exact determination of asymptotic CMB temperature-redshift relation} 
\end{center}
\vspace{0.5cm}
\begin{center}\large{Steffen Hahn}
\end{center}
\vspace{0.1cm}
\begin{center}
{\em Karlsruhe Institute of Technology (KIT), Germany;\\ steffen.t.hahn@gmail.com}
\end{center}
\vspace{1.0cm}
\begin{center}
\large{Ralf Hofmann}
\end{center}
\vspace{0.1cm}
\begin{center}
{\em Institut f\"ur Theoretische Physik, Universit\"at Heidelberg,\\ 
Philosophenweg 16, D-69120 Heidelberg, Germany;\\ r.hofmann@thphys.uni-heidelberg.de
}
\end{center}

\begin{abstract}
Based on energy conservation in a Friedmann-Lemaitre-Robertson-Walker (FLRW) Universe, on the Legendre transformation between energy density and pressure, and on nonperturbative asymptotic freedom at high temperatures we derive the coefficient $\nu_{\rm CMB}$ in the high-temperature ($T$) -- redshift ($z$) relation, $T/T_0=\nu_{\rm CMB}(z+1)$, of the Cosmic Microwave Background (CMB).
Theoretically, our calculation relies on a deconfining SU(2) rather than a U(1) photon gas.
We prove that $\nu_{\rm CMB}=\left(1/4\right)^{1/3}=0.629960(5)$, representing a topological invariant.
Interestingly, the relative deviation of $\nu_{\rm CMB}$ from the critical exponent associated with the correlation length $l$ of the 3D Ising model, $\nu_{\rm Ising}=0.629971(4)$, is less than $2\times 10^{-5}$. We are not yet in a position to establish a rigorous theoretical link between $\nu_{\rm CMB}$ and $\nu_{\rm Ising}$ as suggested by the topological nature of $\nu_{\rm CMB}$ and the fact that both theories share a universality class. We do, however, line out a somewhat speculative 
map from the physical Ising temperature $\theta$ to a fictitious SU(2) Yang-Mills temperature $T$, the latter continuing 
the asymptotic behavior of the scale factor $a$ on $T/T_0$ for $T/T_0\gg 1$ down to $T=0$, and an exponential map from $a$ to $l$ to reproduce critical Ising behavior.   
\end{abstract}
  
\end{titlepage}

That an SU(2) gauge principle underlies the description of thermal photon gases is a theoretically supported  \cite{RH2016} and phenomenologically plausible possibility \cite{RH2009,RH2013,RH2015,HH2017,HKH2017}. In particular implications of the high-redshift relation,  
$T/T_0=\nu_{\rm CMB}(z+1)$, with $\nu_{\rm CMB}=0.63$ (a result obtained by numerical analysis, see below) 
were investigated in view of the redshift $z_{\rm re}$ of (instantaneous) reionization of the 
Universe and the present value of the cosmological expansion rate $H_0$ \cite{RH2015,HH2017}. In both cases discrepancies are resolved between local, that is, low-$z$ cosmological observation on one hand and fits to CMB angular power spectra, including high-$z$ information, on the other hand. Aside from this cosmological motivation, a theoretical understanding of the coefficient $\nu_{\rm CMB}$ is interesting in its own right. Moreover, a venue to {\sl analyse} critical behaviour may emerge for models that share the universality class of SU(2) Yang-Mills thermodynamics concerning its deconfining-preconfining phase transition. The purpose of this short note is to actually {\sl derive} the value of $\nu_{\rm CMB}$.     
      
The underlying equation to determine the $T$-$z$ relation of deconfining SU(2)$_{\rm CMB}$ thermodynamics is \cite{RH2015}
\eqb
\label{enecons}
\frac{\mbox{d}\rho}{\mbox{d}a}=-\frac{3}{a}\left(\rho+P\right)\,,
\eqe
where $\rho$, $P$, and $a$ denote SU(2)$_{\rm CMB}$ energy density and pressure, respectively, and $a$ refers to 
the cosmological scale factor, normalized such that today $a(T_0)=1$, $T_0=2.725\,$K indicating the present baseline temperature of the CMB \cite{COBE}. As usual, $z$ and $a$ are related as $a^{-1}=z+1$. 
Formally, Eq.\,(\ref{enecons}) is solved by 
\begin{equation}\label{formsol}
a=\exp\left(-\frac{1}{3}\int^{\rho(T)}_{\rho(T_0)}\frac{\mbox{d}\rho}{\rho+P(\rho)}\right) = \exp\left(-\frac{1}{3}\int^T_{T_0}\mbox{d}T^\prime\,\frac{\frac{\mbox{d}\rho}{\mbox{d}T^\prime}}{T^\prime\frac{\mbox{d}P}{\mbox{d}T^\prime}}\right) \,.
\end{equation}
To further process the integrand in Eq.\,(\ref{formsol}) it is useful to appeal to the Legendre transformation and its $T$-differentiated version
\eqb
\label{LT}
\rho=T\frac{\mbox{d}P}{\mbox{d}T}-P\,,\ \ \ \ \frac{\mbox{d}\rho}{\mbox{d}T}=T\frac{\mbox{d}^2P}{\mbox{d}T^2}\,.
\eqe
Substituting Eqs.\,(\ref{LT}) into Eq.\,(\ref{formsol}) yields
\begin{equation}\label{sol}
a = \exp\left(-\frac{1}{3}\int^T_{T_0} \mbox{d}T^\prime \, \frac{\mbox{d}}{\mbox{d}T^\prime} \left[\log \frac{s(T^\prime)}{M^3} \right]\right) = \exp\left(-\frac{1}{3}\log\frac{s(T)}{s(T_0)}\right)\,,
\end{equation}
where $M$ denotes an arbitrary mass scale, and the entropy density $s$ is given as 
\eqb
\label{entropydens}
s=\frac{\rho+P}{T}\,.
\eqe
Note that a potential, {\sl direct} contribution to $s$ from the thermal ground state \cite{RH2016} actually is 
absent due to its cancellation in the sum $\rho+P$. The important feature of Eq.\,(\ref{sol}) is 
that its right-hand side solely refers to the boundary values of the state variable $s$ at $T_0$ and $T$. Therefore, this right-hand side is independent of any intermediate thermodynamics dynamics. In this sense, the solution to Eq.\,(\ref{enecons}) of a 
topological nature.

Let us now discuss the situation at high temperatures. For $T\gg T_0$, where the Stefan-Boltzmann limit is well saturated \cite{RH2016}, $s(T)$ is proportional to $T^3$. On the other hand, we know that at 
$T_0$ the excitations of the theory represent a free photon gas because the mass of vector modes diverges. Therefore, $s(T_0)$ is proportional to $T_0^3$. As a consequence, the ratio $s(T)/s(T_0)$ in Eq.\,(\ref{sol}) reads
\eqb
\label{ratentr}
\frac{s(T)}{s(T_0)}=\frac{g(T)}{g(T_0)} \left(\frac{T}{T_0}\right)^3=\left(\left(\frac{g(T)}{g(T_0)}\right)^{\frac{1}{3}}\frac{T}{T_0}\right)^3\,,\ \ \ (T\gg T_0)\,,
\eqe
where $g$ refers to the number of relativistic degrees of freedom at the respective 
temperatures. We have $g(T)=2\times 1+3\times 2=8$ (two photon polarizations plus three polarizations for each of the two vector modes) and $g(T_0)=2\times 1$ (two photon polarizations). Substituting this into Eq.\,(\ref{ratentr}) 
and inserting the result into Eq.\,(\ref{sol}), we arrive at 
\begin{equation}\label{solt>t0}
a = \frac{1}{z+1}=\exp\left(-\frac{1}{3}\log \left(4^{\frac13}\frac{T}{T_0}\right)^3\right) = \left(\frac14\right)^{\frac13}\frac{T_0}{T}\equiv \nu_{\rm CMB}\frac{T_0}{T} \,,\ \ \ (T\gg T_0)\,,
\end{equation}
proving our above claim. In \cite{HH2017} we have used the 
value $\nu_{\rm CMB}=0.63$ which is a good approximation to 
\eqb
\nu_{\rm CMB}=\left(\frac14\right)^{\frac13}=0.629960(5)\,.
\eqe
In \cite{YaffeSvet} it was argued that the order parameter for deconfinement in 
4D SU(2) Yang-Mills thermodynamics, the 3D Polyakov-loop variable ${\cal P}$, should obey 
long-range dynamics governed by (electric) ${\bf Z}_2$ center symmetry which is the global symmetry of 
the Ising model void of an external magnetic field. Sharing a universality class, the critical exponents 
of the correlation lengths thus should coincide for both theories. The superb agreement between 
$\nu_{\rm Ising}=0.629971(4)$ \cite{Kos2016} and $\nu_{\rm CMB}$, 
\eqb       
\left|\frac{\nu_{\rm CMB}-\nu_{\rm Ising}}{\nu_{\rm Ising}}\right|\sim 1.7\times 10^{-5}\,,
\eqe
is indicative of such a link between the two theories even though $\nu_{\rm CMB}$ appears as a 
coefficient in a cosmological $T$-$z$ relation and {\sl not} as a critical exponent 
governing the divergence of the  correlation length $l$. Still, $\nu_{\rm CMB}$ is a topologically determined 
number, a feat being characteristic of a critical exponent. We emphasize that an analytic derivation 
of a 3D Ising exponent via 4D SU(2) Yang-Mills thermodynamics is an extremely useful benchmark. However, in spite of the 
above arguments a rigorous link is not in sight presently. 

\begin{figure}
\centering
\includegraphics[width=0.5\textwidth]{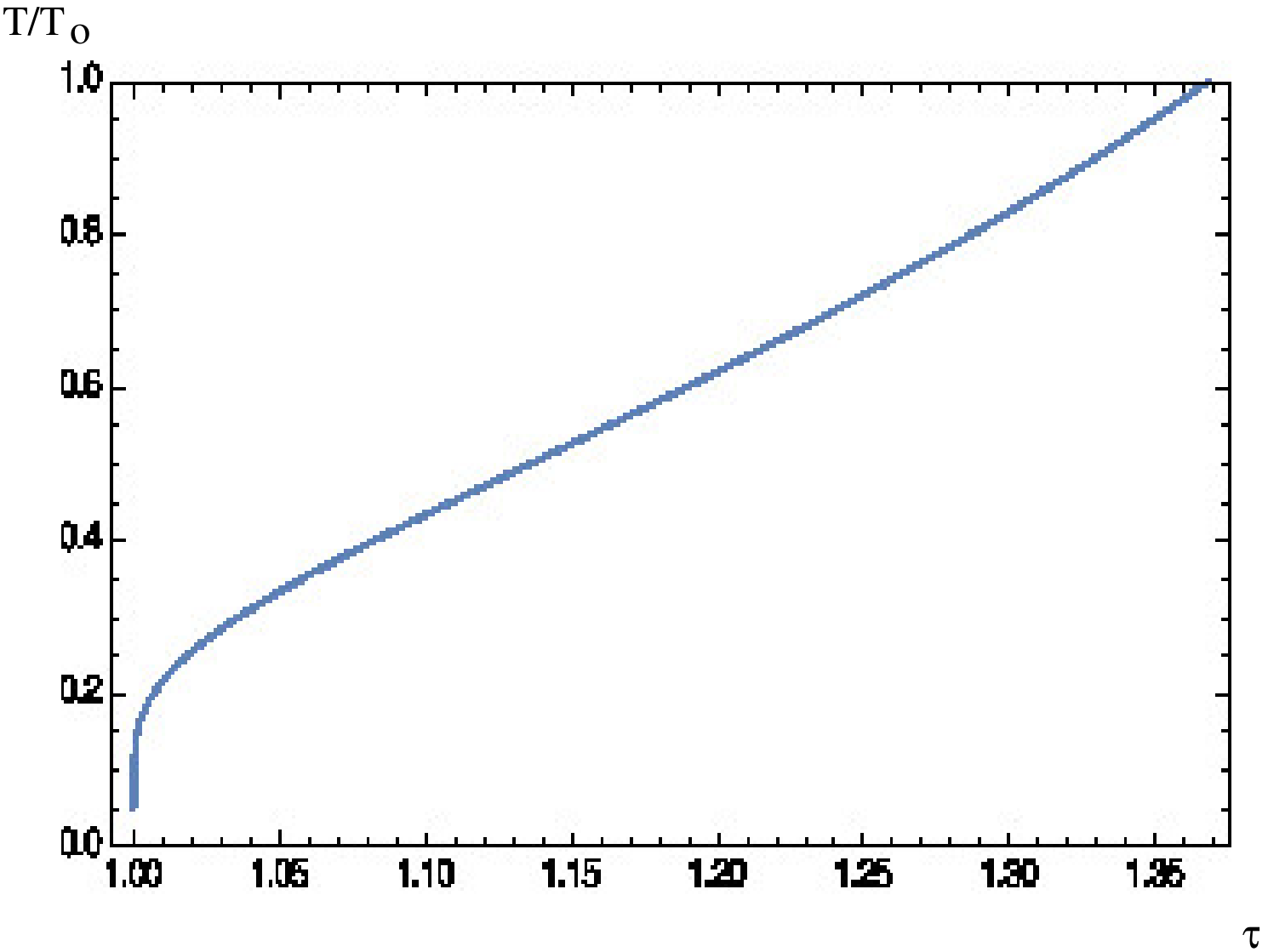}
\caption{\protect{\label{Fig-1}} Plot of the function $\frac{T}{T_0}(\tau)$ in Eq.\,(\ref{IsingYM}).}      
\end{figure}

What one can try to do though, is to continue the asymptotic behavior of Eq.\,(\ref{solt>t0}) down 
to $T=0$, where $a$ diverges, positing that a monotonic increasing function of $a$ describes 
the ratio $l/l_0$ of the Ising correlation length $l$ to some reference length $l_0$ and that 
therefore already the Yang-Mills asymptotics catches the downward approach of scale invariance 
at the second order transition of the Ising model. This renders $T$ a 
fictitious temperature in deconfining SU(2) Yang-Mills theory (for $T\le T_0$ 
this theory does not obey the asymptotics of Eq.\,(\ref{solt>t0}) \cite{RH2016}). One may now 
construct a map from the physical 
Ising temperature $\theta$ to $T/T_0$ as 
\eqb
\label{IsingYM}
\frac{T}{T_0}=-\frac{1}{\log(\tau-1)}\,,
\eqe
where $\tau\equiv\frac{\theta}{\theta_c}$, and $\theta_c$ denotes the critical temperature of the Ising phase transition.  
Obviously, $T\searrow 0$ implies that $\tau\searrow 1$, and $T/T_0\searrow 1$ associates 
with $\tau\searrow 1+\e^{-1}\sim 1.367879$, see also Fig.\,\ref{Fig-1}. 
Substituting Eq.\,(\ref{IsingYM}) into Eq.\,(\ref{solt>t0}) and exponentiating, we arrive 
at
\eqb
\label{Isingcl}
\exp(a)=(\tau-1)^{-\nu_{\rm CMB}}\,.
\eqe 
As $\tau\searrow 1$ this yields the same $l/l_0$ critical behavior for $\exp(a)$, and 
the above-mentioned monotonic function thus turns out to be the exponential map.

\end{document}